%% file: 0-Main.tex
\documentclass[11pt]{article}
\usepackage{cite}
\usepackage{amsmath,amssymb,amsfonts}
\usepackage{algorithmic}
\usepackage{url,graphicx,multirow}
\usepackage{textcomp}
\usepackage{longtable}
\usepackage[utf8]{inputenc}
\usepackage{lipsum}
\usepackage{listings}
\usepackage{placeins}
\usepackage{float}
\usepackage{array}

\graphicspath{ {figures/}}
\usepackage{parskip}

\begin{document}

\title{Systematic Mapping Protocol: Reasoning Algorithms on Feature Model\\
\vspace{15mm}
\footnotesize{Final version} }
\author{}

\maketitle
\begin{center}
    Dr. Samuel Sepúlveda\footnote{
samuel.sepulveda[at]ufrontera.cl\\Dpto. Cs. de la Computación e Informática\\Centro de Estudios en Ingeniería de Software\\
Universidad de La Frontera\\ Temuco, Chile} and Mg. Marcelo Esperguel\footnote{marcelo.esperguel[at]uautonoma.cl\\Universidad Autónoma de Chile\\ Temuco, Chile}
\end{center}

\newpage

\textit{\textbf{Abstract} }

\textbf{Context:} The importance of the \textit{feature modeling} for the \textit{software product lines} considering the modeling and management of the variability.\\
\textbf{Objective:} Define a protocol to conduct a \textit{systematic mapping study} to summarize and synthesize the evidence on \textit{reasoning algorithms} for \textit{feature modeling}.\\
\textbf{Method:} Application the protocol to conduct a \textit{systematic mapping study} according the guidelines of K. Petersen.\\
\textbf{Results:} A validated protocol to conduct a \textit{systematic mapping study}.\\
\textbf{Conclusions:} Initial findings show that a more detailed review for the different \textit{reasoning algorithms} for \textit{feature modeling} is needed. \\



\textbf{Keywords:} \textit{Reasoning algorithm, Feature model, Software product lines, systematic mapping}.

\newpage
\input{1-Introduction}

\input{2-Definition_protocol}

\input{3-Conclusion}

\section*{Acknowledgments}

Samuel Sepúlveda thanks to Vicerrectoría de Investigación y Postgrado, Universidad de La Frontera, research project DI20-0014. Thanks to Jonathan Jara and Sebastián Pardo for their useful technical support.

\bibliographystyle{bib/IEEEtran}

\bibliography{bib/0-Main.bib}

\end{document}

%% file: 1-Introduction.tex
\section{Introduction}
\label{intro}

This paper aims to present the protocol definition to carry out systematic mapping over the last ten years about reasoning algorithms in feature models (FM), used in the stages and activities that compose the software product line framework. 

The idea behind considering only works from the last 10 years is due to the fact that a similar report was published in 2010 \cite{benavides2010automated20years} which considers very similar aspects. Therefore, this proposal would only consider works posterior to the publication of Benavides. 

The systematic mapping will be the beginning of the DIUFRO DI20-0014 research project that aims to evaluate the development and implementation of reasoning algorithms based on the modeling driven development (MDD) approach. \par

This report is structured as follows. The next section section describes the research method conducted, including the definition of search string, search process and the criterion for including and excluding papers. Finally, section \ref{conclusion} presents the conclusions and future work.

%



%% file: 2-Definition_protocol.tex
\section{Protocol definition}
\label{def_protocolo}
This section describes the protocol definition to conduct a systematic mapping study (SMS) according the guidelines defined by Petersen \cite{petersen2015guidelines}.

\subsection{Objective}
\label{objetivo}

Collect reasoning algorithms proposals for feature models, present in the literature of the last 10 years, with the intention of collecting the largest amount of data from the proposals found for their analysis, synthesis and subsequent publication of the results generated.

\subsection{Need}
\label{necesidad}

In order to generate a proposal for reasoning algorithms based on MDD approaches is necessary to know in detail the proposals that exist today within the area, as this will allow:

\renewcommand{\labelenumi}{\alph{enumi})}
 \begin{enumerate}
    \item To know the necessary requirements to be able to create algorithms of this nature.
    \item Know what technologies, tools, approaches or others are used in the creation of these algorithms and understand the justifications for use in each case.
    \item Avoiding activities or processes already carried out by other authors.
    \item Avoid designing or creating processes/assets that have already been exposed by other authors.
 \end{enumerate}
 
However, the last report/study that collects this information is 10 years old \cite{benavides2010automated20years} and while similar and more recent state of the art reports have emerged \cite{benavides2013automatedFM&CM, galindo2019automatedQuoVadis}, these have a different focus than the one we wish to address in this paper, in particular \cite{benavides2013automatedFM&CM} is an extension of \cite{benavides2010automated20years} and seeks to answer questions related to FM reasoning algorithms that can be applied in \textit{configuration modeling}, and on the other hand Galindo presents results focused mostly on bibliometrics \cite{galindo2019automatedQuoVadis}. 

It is expected that the publication of the results will not only serve as an input for the following stages of the research project or for the formulation of future works, but it will also serve the research community in SPL and FM as a synthesis of what happened in the last 10 years in reasoning algorithms, both at the theoretical and bibliometric levels.

\subsection{Research questions}
\label{rqs}

According to Kitchenham and Charters \cite{Kitchenham07guidelinesfor}, we define a context for the research questions (RQ) guiding this study. The context for the RQs arises from the hypotheses of this project and from a more general question: \textit{Will it be possible to build a set of reasoning algorithm based on a modeling language composed by a metamodel and OCL constraints for feature models in SPL context ?}. 

To answer this question, it is necessary to know the existing proposals in the literature related to reasoning algorithms, this information will be useful to also know the technologies used and the context in which these algorithms has been used. 

The table \ref{tab:rq_table} shows each RQ defined for the protocol, the table also shows a possible classification and sub-questions derived from this, finally specifies the objective that the RQ seeks to clarify.

%
%

\begin{table}[hbt!]
\begin{center}
\begin{tabular}{ | p{0.5cm} | p{3.5cm}| p{3.5cm} | p{4cm} |} 
\hline
{\footnotesize ID}& {\footnotesize RQ} &{\footnotesize Classification} & {\footnotesize Objective} \\ 
\hline
{\footnotesize RQ1 }& {\footnotesize  In what stages of SPL are these algorithms used?}
& 
 \renewcommand{\labelitemi}{\textperiodcentered}
  \footnotesize - Domain Engineering  \par
  \footnotesize - Application Engineering  \par
  {\footnotesize Sub Classification:} \par
{\footnotesize RQ1.1 In what process of the stage is it used? } \par
{\footnotesize RQ1.2 What is the purpose of the algorithm?}
&
{\footnotesize Highlight areas where algorithms are applied and for what purpose.
Highlight areas that have more/less research.}
\\ 
\hline
{\footnotesize RQ2}& {\footnotesize What type of technologies do algorithms mainly use?} 
&
 \renewcommand{\labelitemi}{\textperiodcentered}

  {\footnotesize Meta-model, UML, OCL, Transformations, Solver , others.} 

&
{\footnotesize To know the most used technologies that the algorithms are based.
Highlight the different existent possibilities for researchers to implement new algorithms.} \\
\hline
{\footnotesize RQ3}& {\footnotesize Which is the origin of the proposal? }
&
\renewcommand{\labelitemi}{\textperiodcentered}

 - {\footnotesize Academy.} \par
 - {\footnotesize Industry.}\par
 - {\footnotesize Both.}\par

&
{\footnotesize } \\
\hline
{\footnotesize RQ4}& {\footnotesize Which is the level of validation?}
&
\renewcommand{\labelitemi}{\textperiodcentered}

 \footnotesize Wieringa’s classification \cite{wieringa2006requirements} 

&
{\footnotesize Gain insight into the research maturity level based on the Wieringa research taxonomy.
} \\
\hline
{\footnotesize RQ5}& 
{\footnotesize What kind of FM does the algorithm work on?}
&
\renewcommand{\labelitemi}{\textperiodcentered}

\footnotesize {Extended FM, Multiplicity, Orthogonal Model, Multi FM, Complex FM, others.}

&
{\footnotesize Highlight the most used FM when creating algorithms.
To know what type of models each algorithm works on.} \\
\hline
{\footnotesize RQ6}
& 
{\footnotesize What problems does the algorithm solves?}
&
\renewcommand{\labelitemi}{\textperiodcentered}

\footnotesize Null FMs.
\footnotesize Valid product.
\footnotesize Valid partial configuration.

&
{\footnotesize Highlight which problems have more solutions and which don't.} \\
\hline

\end{tabular}
\end{center}
\caption{RQ and details.}
\label{tab:rq_table}
\end{table}

\subsection{Publication Questions}

Additionally, a set of publication questions (PQs) has been included to complement the gathered information and characterize the bibliographic and demographic space. This includes the type of venue where the papers were published, and the amount of papers per year, details are shown in Table \ref{tab:pq_table}.

%
%

\begin{table}[hbt!]
\begin{center}
\begin{tabular}{ | p{0.8cm} | p{3.5cm}| p{2.5cm} | p{4cm} |} 
\hline
{\footnotesize ID} & {\footnotesize PQ}
&
{\footnotesize Classifications}
&
{\footnotesize Objective} \\

\hline
{\footnotesize PQ 1} 
&
{\footnotesize Place of publication of the article}
&
\renewcommand{\labelitemi}{\textperiodcentered}

- {\footnotesize Journal.} \par
- {\footnotesize Congress.}

&
{\footnotesize Help researchers know which are the journals/congress that are most interested in each topic and to which they can point to send new works. (Quo Vadis).} \\
\hline
{\footnotesize PQ 2} 
&
{\footnotesize Which is the publication year of the article?}
&

{\footnotesize 2010-2020.}

&
{\footnotesize Highlight how the algorithms have progressed through the years.
Years with most publications.}\\
\hline
\end{tabular}
\end{center}
\caption{PQ and details.}
\label{tab:pq_table}
\end{table}

%
%

\subsection{Data Sources}

According to \cite{Kitchenham07guidelinesfor, brereton2007lessons} we consider the data sources detailed in Table \ref{tab:ds_table}, that are recognized among the most relevant in the SE community.

\begin{table}[hbt!]
\begin{center}
\begin{tabular}{ | p{4cm} | p{8cm} |} 
\hline

{\footnotesize Library} & {\footnotesize Url} \\

\hline
{\footnotesize ACM Digital Library} & {\footnotesize https://dl.acm.org} \\

\hline
{\footnotesize IEEE Xplore} & {\footnotesize https://ieeexplore.ieee.org/Xplore/home.jsp} \\

\hline
{\footnotesize Science Direct} & {\footnotesize https://www.sciencedirect.com} \\

\hline
{\footnotesize Springer Link} & {\footnotesize https://www.springer.com/} \\

\hline
{\footnotesize Wiley Inter-Science} & {\footnotesize https://www.onlinelibrary.wiley.com/search/advanced} \\

\hline
\end{tabular}
\end{center}
\caption{Data Sources}
\label{tab:ds_table}
\end{table}

\subsection{Search String}

The search string has been constructed according to the steps defined in \cite{Kitchenham07guidelinesfor}, that is, from the context and research question have been extracted a set of keywords, then for each keywords has beend proposed a set of synonims. Finally using PICOC \cite{petticrew2008systematic} the search string is constructed.

The list of keywords and synonyms is listed as follows:

- feature model / modelling
- variability model / modelling
- feature diagram

- reasoning / reasoner
- analysis / analyzer
- automated 
- automated support
- automatic verification
- computer-aided
- automated analysis

- software product family
- software product lines

- algorithm
- solver (?)
- model checking/ validation / verification / querying
- reasoning model

The final query string is described as follows:

{
\small
\begin{figure}[hbt!]
    \centering
    \begin{verbatim}
("feature model” OR "feature models" OR "feature modelling" 
OR "feature diagram" OR "configuration model" 
OR "variability model" OR "variability modelling") 
AND
("reasoning" OR "analysis" OR "analyses") 
AND
("algorithm" OR "automated" OR "computer aided")
AND
("software product line" OR "software product lines"
OR "product family" OR "product families" 
OR "product line" OR "product lines")
    \end{verbatim}
    \caption{Search String}
    \label{fig:search_string}
\end{figure}
}

It is important to mention that the search string has been adapted for some search engines, due to the limitations of each one. However, each adaptation does not add or remove any filter. Table \ref{tab:tr_table} shows the total results for the search string used in the data sources selected.

%
%



\begin{table}[hbt!]
\begin{center}
\begin{tabular}{ | p{4cm} | p{4cm} |} 
\hline
{\footnotesize Library} & {\footnotesize Result} \\
\hline

{\footnotesize ACM} & {\footnotesize 777} \\

\hline

{\footnotesize Springer Link} & {\footnotesize 819} \\

\hline

{\footnotesize IEE Xplore} & {\footnotesize 133} \\

\hline

{\footnotesize Wiley Online Library} & {\footnotesize 35} \\

\hline

{\footnotesize Total Result} & {\footnotesize 1764} \\
\hline
\end{tabular}
\end{center}
\caption{Search string, total results (17th November 2020).}
\label{tab:tr_table}
\end{table}

%
%

\subsection{Search and Selection Process}

The search process starts with an automatic search on selected electronic databases using the search string. The goal of this is to get the first collection of papers to distribute to the team researchers.  Each researcher independently have to filter this collection according to the following criterion, allowing to decide if the initial papers collection are relevant or not for this study, only reviewing title, abstract, and keywords for each papers. 

The first filter is to apply the inclusion criteria (IC) and the remaining papers will be filtered applying the exclusion criteria (EC). The definition of each IC and EC is shown in Tables \ref{tab:ic_table} and \ref{tab:ec_table}.

\begin{table}[hbt!]
\begin{center}
\begin{tabular}{ | p{2cm} | p{10cm} |} 

\hline
{\footnotesize ID} & {\footnotesize Criteria} \\

\hline
{\footnotesize \textbf{CI1}} & {\footnotesize For papers with more than one version, the last version will be included and the others will be excluded. } \\

\hline
{\footnotesize \textbf{CI2}} & {\footnotesize Works written in English.} \\

\hline
{\footnotesize \textbf{CI3}} & {\footnotesize Type of paper:
    \begin{itemize}
     \item Proceeding.
     \item Research Article.
     \item Article.
     \item Conference Paper.
     \item Journal.
   \end{itemize}} \\

\hline
{\footnotesize \textbf{CI4}} & {\footnotesize Papers published between 2010-2020.} \\

\hline
{\footnotesize \textbf{CI5}} & {\footnotesize Papers whose abstracts show the relationship with the automatic analysis of Feature Models.} \\

\hline
{\footnotesize \textbf{CI6}} & {\footnotesize Topic:
    \begin{itemize}
     \item Computer Science.
   \end{itemize}} \\

\hline
\end{tabular}
\end{center}
\caption{Inclusion criteria}
\label{tab:ic_table}
\end{table}

\begin{table}[hbt!]
\begin{center}
\begin{tabular}{ | p{2cm} | p{10cm} |} 

\hline
{\footnotesize ID} & {\footnotesize Criteria} \\

\hline
{\footnotesize \textbf{CE1}} & {\footnotesize Duplicated papers will be excluded} \\
\hline
{\footnotesize \textbf{CE2}} & {\footnotesize The following types of papers will be excluded:
    \begin{itemize}
     \item Tutorials.
     \item Short Paper.
     \item Abstract.
     \item Poster.
     \item Keynote.
     \item Paper in progress (incomplete).
     \item Books.
     \item Book Chapter.
   \end{itemize}} \\
\hline
{\footnotesize \textbf{CE3}} & {\footnotesize Papers whose abstracts doesn't show the relationship with the automatic analysis of Feature Models.
} \\
\hline
{\footnotesize \textbf{CE4}} & {\footnotesize Secondary researches. (If they exist, they will be added as related work).} \\
\hline
{\footnotesize \textbf{CE5}} & {\footnotesize Papers that cannot be accessed.} \\
\hline
{\footnotesize \textbf{CE6}} & {\footnotesize Papers published before 2010.} \\
\hline
{\footnotesize \textbf{CE7}} & {\footnotesize Articles with an extension of less than 4 pages.} \\
\hline
\end{tabular}
\end{center}
\caption{Exclusion criteria}
\label{tab:ec_table}
\end{table}

\subsection{Resolving differences and avoiding bias}

In order to reduce the differences as much as possible, our protocol declares the following decisions

\begin{itemize}
  \item Collaborative definition of the SLR protocol and RQs.
  \item External validation for the search string
  \item Publication of the protocol for public scrutiny on the arXiv platform

  \item To avoid any potential bias due to a particular researcher examining each paper, we verified that the manner of applying and understanding the exclusion criteria was similar for the researchers involved in the SMS (inter-rater agreement).
  \item The researchers individually decide on the inclusion orexclusion of a sub set of assigned papers randomly chosen from those retrieved by a pilot selection.
  \item A test of concordance based on the Fleiss’ Kappa statistic will be performed as a means of validation \cite{gwet2002kappa}. If \textit{Kappa $\geq 0.75$} then the criteria is clear enough \cite{fleiss2013statistical}, else, the criterion must be reviewed to get by its interpretation and application.
\end{itemize}

Another way to do this task is by following the criteria defined by \cite{petersen2015guidelines}.

\subsection{Results and Reporting}

Finally, for the selected papers, relevant data will be extracted to answer the RQs and PQs.

The meta-data collected for each paper: (i) title, (ii) authors (each of them), (iii) publication year, (iv) type of publication and ranking, (v) algorithm reported (vi) technology based reported, (vii)  and (viii) results and future work.

\begin{figure}[h]
    \centering
    \includegraphics[scale=0.49]{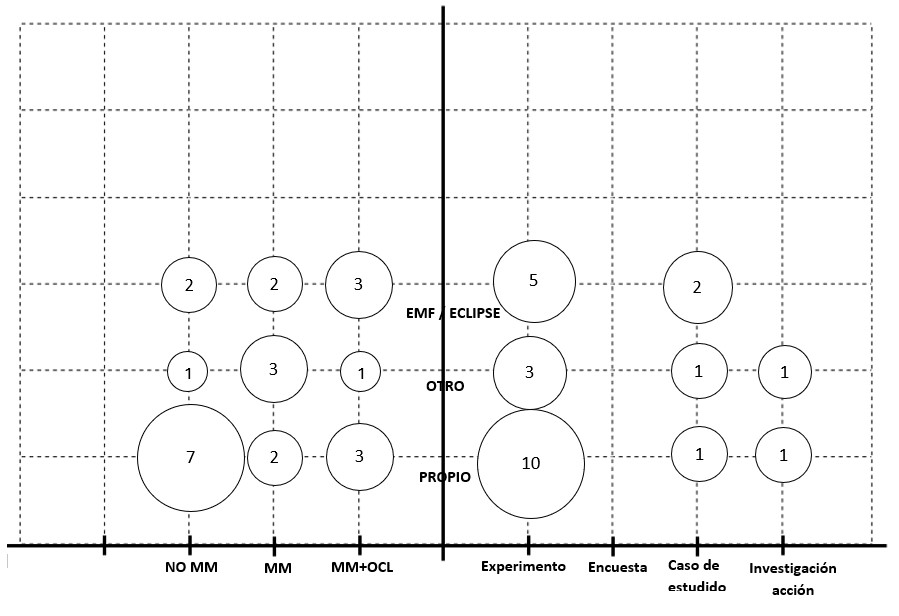}
    \caption{Bubble Diagram Example}
    \label{fig:bubble_diagram}
\end{figure}

Another result will be a bubble diagram that graphically represents the number of papers found divided into categories. An example is represented in the figure \ref{fig:bubble_diagram}.

Presenting the results of this SMS considers 2 stages:
\begin{itemize}
    \item stage 1: publish the SMS protocol at Arxiv web platform\footnote{https://arxiv.org},
    \item stage 2: publish the SMS results in an academic journal or conference focused on the topic of FM and SPL.
\end{itemize}

The structure recommended by \cite{Kitchenham07guidelinesfor} will be used to report the results for stage 2. The main sections are:
\begin{itemize}
    \item Introduction
    \begin{itemize}
        \item Context
        \item Motivation
        \item Aim and need
    \end{itemize}
    \item Background
    \begin{itemize}
        \item SPL
        \item Variability
        \item Feature modeling
        \item Reasoning algorithms
    \end{itemize}
    \item Methodology
    \item Results and discussion
    \begin{itemize}
        \item Answers to RQs
        \item Threats to validity
    \end{itemize}
    \item Related work
    \item Conclusion
    \begin{itemize}
        \item Conclusions
        \item Future work
    \end{itemize}
\end{itemize}

%% file: 3-Conclusion.tex
\section{Conclusion and Future Work}
\label{conclusion}
We presented a protocol definition of a SMS to summarize and synthesize the evidence about the automation on feature modeling reasoning algorithms.

The initial results show that we need to do a more detailed review for the different kind of reasoning algorithms for feature modeling and their impact on an eventual proposal and prototype tool support.

As a future work we plan to define a set of FM reasoning algorithms improving performance over large FMs and streamline variability management in SPL.